\documentclass[11pt]{article}

\usepackage{color,graphicx}
\usepackage{young}
\usepackage[vcentermath]{youngtab}
\usepackage{amsmath,amssymb,graphicx}
\usepackage{hyperref}
\definecolor{darkred}{rgb}{0.65,0.15,0}
\hypersetup{pdfborder={0 0 0},colorlinks=true,urlcolor=darkred,citecolor=blue,linkcolor=darkred,linktocpage=true}

\usepackage{cite}
\usepackage{amsmath}
\usepackage{amsfonts}
\usepackage{amssymb}
\usepackage{graphicx}%
\usepackage{amsthm}
\usepackage{mathrsfs}
\usepackage[T1]{fontenc}
\usepackage{enumerate}
\usepackage{color}

\def\4diml{four-dimensional}

\def\-1{^{-1}}

\makeatletter

\@addtoreset{equation}{section}
\makeatother

\begin{document}

\thispagestyle{empty}

\begin{center}
\vspace{40mm}
{\LARGE \bf Cosmological solutions in the Brans-Dicke theory
\\[2mm] via invariants of symmetry groups }

\vspace{12mm}
\normalsize
{\large   E. Ahmadi-Azar\footnote{e.ahmadi.azar@azaruniv.ac.ir},
K. Atazadeh\footnote{atazadeh@azaruniv.ac.ir}, A. Eghbali\footnote{eghbali978@gmail.com},}

\vspace{2mm}
{\small \em Department of Physics, Faculty of Basic Sciences,\\
Azarbaijan Shahid Madani University, 53714-161, Tabriz, Iran}\\

\vspace{7mm}

%\hrule

\vspace{6mm}

\begin{tabular}{p{15cm}}
{\small
We proceed to obtain an exact analytical solution of the Brans-Dicke (BD) equations for the spatially flat
($k=0$) Friedmann-Lamaitre-Robertson-Walker (FLRW) cosmological model in both cases of the absence and presence of the cosmological constant.
The solution method that we use to
solve the field equations of the BD equations is called the ``invariants of symmetry groups method'' (ISG-method).
This method is based on the extended Prelle-Singer (PS) method and it employs the Lie point symmetry, $\lambda$-symmetry,
and Darboux polynomials (DPs). Indeed, the ISG-method tries to provide two independent first-order
invariants associated to the one-parameter Lie groups of transformations keeping ordinary differential equations (ODEs) invariant, as solutions.
It should be noted for integrable ODEs, the ISG-method guarantees the extraction of these two invariants.
In this work for the BD equations in FLRW cosmological model, we find the Lie point symmetries, $\lambda$-symmetries
and DPs, and obtain the basic quantities of the extended PS method (which are the null forms and the integrating factors).
By making use of the extended PS method we find two independent first-order invariants, in such a way appropriate
cosmological solutions from solving these invariants as a system of algebraic equations are simultaneously obtained.
These solutions are wealthy so that they include many known special solutions,
such as O'Hanlon-Tupper vacuum solutions, Nariai's solutions, Brans-Dicke dust solutions, inflationary solutions, and etc. }

\end{tabular}
\vspace{-1mm}
%\hrule
\end{center}

\setcounter{page}{1}

\tableofcontents

%%%%%%%%%%%%%%%%%%%%%%%%%%%%%%%%%%%%%%%%%%%%%%%%%%%%%%%%%%%%%%%%%%%%%%%%%%%%%%%%%%%%%
\section{\label{Sec.I} Introduction}

Mach's ideas strongly inspired Albert Einstein, so he attempted to include this principle in his theory
when formulated the theory of general relativity (GR).
During the years 1907-1916 when Einstein was developing his theory of GR, he strongly influenced by Mach's philosophy and tried to include this principle in his theory in every possible.
Finally, he could reflect the relationship between the geometry of space-time and also the distribution of matter in the GR equations.
Later, it was shown that the Einstein's field equations (EFEs) can be include solutions in the absence of matter and in empty space (Minkowski space-time, Taub-NUT space-time, etc.) \cite{ref.1}.
Therefore, contrary to Mach's opinion (the stronger form of Mach's principle), the GR theory includes the weaker form of Mach's principle, and hence space-time
has an existence independent of matter.
Among physicists, there is a group that accepts Mach's opinion and believes that GR, which does not include the stronger form of Mach's principle, should be modified to include it.
Since the 70s, efforts have been made in this direction, among which we can mention the scalar-tensor theories (STTs) \cite{ref.2}.
The Brans-Dicke (BD) theory of gravity (sometimes called the Jordan-Brans-Dicke scalar-tenser theory of gravity) is the simplest version of the STT,
which was developed by Pascual Jordan in the 1950s and later by Brans and Dicke in the early 1960s \cite{{ref.3},{ref.4}}.
Actually, theories about this had already been proposed years earlier by Jordan, Fierz and Thiery.
One of the motivations of these authors was to include the stronger form of the Mach's principle in gravity. This principle
states that the inertial mass of a body is due to all the rest of the matter in the universe.
For the first time, the idea of varying the Newton's gravitational constant $G$ with the age of the universe was proposed by Dirac \cite{ref.5}.
His purpose in doing this was to explain the weakness of the gravitational force among the fundamental forces in nature.
In the Jordan gravity theory, the variation of the constant $G$ with space-time
was described by a simple scalar field.
However, scientists such as Fierz and Bondi criticized this theory, because energy was not conserved in it.
In 1961, Brans and Dicke proposed \cite{ref.3} a more complete theory of gravity incorporating a varying gravitational constant (see also \cite{ref.4}).
In the BD theory, the constant $G$ has been replaced by the inverse of a scalar field $\varphi$.
The scalar field $\varphi$ plays the role of the $G$, and Mach's principle is revived through it.
Also, in the BD theory a dimensionless parameter $\omega$, which measures the coupling of the $\varphi$ to matter, is included.
In the GR, there is only one tensor field $g_{\mu \nu}$ describing the geometry of space-time, while in the BD theory,
in addition to the tensor field $g_{\mu \nu}$, a scalar field $\varphi$ is incorporated for reviving Mach's principle.
In 1980, Batista \cite{ref.6} extended the cosmological model obtained by Brans and Dicke to non-homogeneous models.
He considered density $\rho$ and pressure $p$ as functions of the variables $r, t$ and showed that for the dust-dominant universe,
the same solutions for $a(t)$ and $\varphi$ are obtained as previously found by Brans and Dicke.
In other words, the assumption of inhomogeneity of dust distribution goes not change the BD solutions.
McIntosh \cite{ref.7} showed that with the spatially flat FLRW metric, the scalar field $\varphi$, the pressure $p$ and its density $\rho$,
there are solutions for the BD equations. He showed that for solutions of the BD field equations, it must not be assumed that the $\varphi$, $p$ and $\rho$ are only functions of the variables
on which the metric $g_{\mu \nu}$ depend. This true not just in the cosmological case but in vacuum and other cases as well.
The vacuum solutions of BD equations for the spatially flat FLRW metric were obtained by O'Hanlon and Tupper \cite{ref.8}, Chauvet \cite{ref.9}, Cervero \cite{ref.10} and Lorentz-Petzold \cite{ref.11}.
Chauvet also obtained the solutions of the BD equations parametrically in the case $k=\pm 1$.
Uehara and Kim \cite{ref.12} studied the BD theory in the presence of the cosmological constant $\Lambda$, in such a way that they obtained solutions for the
BD equations in the presence of the cosmological constant for the spatially flat FLRW metric.
They also obtained an exponential solution for the radiation universe $(p=\rho/3)$ when $\omega =-3/2$.

Cosmological dynamical equations (DEs) of FLRW model in the framework of GR theory and many other models in the framework of extended gravity theories such
as BD theory \cite{ref.13}, $f(T)$ gravity \cite{{ref.14},{ref.14.1},{ref.14.2}}, $f(R)$ theory \cite{{ref.15},{ref.15.1},{ref.15.2}}, Rastall theory \cite{ref.16}, etc., have been solved analytically and numerically by different methods.
But, there is no unique solution method that is general, so that it can be used not only to solve the field equations of BD theory, but also the field equations of any extended gravitational theories.
Note that the cosmological DEs are coupled non-linear differential equations, which require special methods.
The PS-method, which was later modified and completed by Chandrasekar {\it et al.} \cite{ref.17}, is one of these methods.
This solution method has made great progress in the field of theory of non-linear differential equations.
The ability that this method has shown in solving a number of well-known non-linear differential equations, such as the $2$-dimensional Kepler problem,
the modified Emden equation, and the Helmholtz oscillator \cite{ref.18}, is indicative of the fact that the
modified extended PS-method, when combined with other symmetry methods, it can be a powerful algorithmic
solution for solving non-linear DEs in physics, such as cosmology \cite{{ref.19},{ref.20}}.
We may use this method to extract independent invariants from the cosmological DEs in a mini-super space, and hence we have actually solved the problem,
because these invariants can be considered as a system of algebraic equations.
Solving this system of algebraic equations is much easier than solving the system of cosmological DEs (which are in the form of non-linear differential equations).
It should be noted that the extended PS-method and combined with other symmetry methods that we call the ISG-method, they are not included in other branches of science, especially gravity and cosmology.
The remarkable point in this method is that it is difficult to calculate its basic quantities, i.e.,
null form $S$ and integrating factor $R$ in the extended PS-method are related to the fundamental quantities of other symmetry methods such as the infinitesimals $\tau$ and $\xi$ in the Lie point symmetry and the function $\lambda$ in the $\lambda$-symmetries, and DPs in the DPs method \cite{{ref.19},{ref.20},{ref.21},{ref.22},{ref.23},{ref.24},{ref.25},{ref.26}}.
Recently, we have presented the ISG-method in order to solve the field equations of the spatially flat FLRW cosmological model
in the presence of $\Lambda$ in the context of GR \cite{ref.27}.
Here, we employ this method to obtain the solutions of the field equations of BD theory for spatially flat FLRW cosmological model
in both cases of the absence and presence of the cosmological constant.
We show that when the coupling constant $\omega$ tends to infinity, the cosmological solutions of GR theory can be obtained from
the analytical solutions of BDEs. For the dust dominated universe, when
the cosmological constant is present, we conclude that the solutions of Uehara and Kim \cite{ref.20} are special cases of the solutions of BDEs.
In addition, we show that the case $\Lambda=0$ of the solutions of BDEs include  Nariai's solutions, BD dust solutions, radiation solutions,
inflationary solutions, O'Hanlon-Tupper vacuum solutions and finally general relativity solutions.

The paper is organized as follows. In Sec. \ref{Sec.II}, we briefly describe the ISG-method applicable for second-order ODEs in physics, especially in cosmology. We give the main points of the ISG-method in this section.
In Sec. \ref{Sec.III}, we first introduce the BD theory of gravity, and then extract the DEs of the FLRW cosmological model in this theory. Then, by combination these equations, we will reduce them to a second-order ODE called the Friedmann-Brans-Dicke equation (FBD-equation). In Sec. \ref{Sec.IV}, we will apply the ISG-method to solve the FBD-equation. This solution enables us to obtain the cosmological solutions of the BD theory.
In Sec. \ref{Sec.V}, the solutions of the BDEs are completed so that they are suitable for cosmological applications.
In Sec. \ref{Sec.VI}, we will examine the cosmological solutions in special cases.
Finally, conclusions are reported in Sec. \ref{Sec.VII}.

%%%%%%%%%%%%%%%%%%%%%%%%%%%%%%%%%%%%%%%%%%%%%%%%%%%%%%%%%%%%%%%%%%%%%%%%%%%%%%%%%%%%%%%%%%%%%%%%%%%%%%%%%
\section{A short review of the ISG-method}
\label{Sec.II}
In order to introduce the ISG-method \cite{ref.27}, let us consider a particle of unit mass moveing in one-dimensional configuration space $\mathbb{Q}=(q)$.
A force ${\bf F}=\phi(t, q,\dot{q}) { \partial}/{\partial q}$ is applied to the particle and gives it an acceleration ${\bf a}=\ddot{q} { \partial}/{\partial q}$.
According to Newton's second law of motion, the DE of the particle is as follows:
\begin{equation}
\ddot{q}=\phi(t, q, \dot{q}), \label{2.1}
\end{equation}
where $\phi$ as a function of  $t$, $q$ and $\dot{q}$  is called a force function of the particle.
Here, ``$^.$'' denotes the total derivative operator with respect to the time $t$, which is defined as
\begin{equation}
^.=\frac{d}{dt} = \frac{\partial}{\partial t} + \dot{q} \frac{\partial}{\partial q} + \phi(t, q, \dot{q})  \frac{\partial}{\partial \dot{q}}. \label{2.2}
\end{equation}
In the force function $\phi(t, q, \dot{q})$, time $t$ is the independent variable, and coordinate $q$ is the dependent variable.
Many physical phenomena have a DE in the form of equation \eqref{2.1}.
That is, the dynamics of such phenomena can be imagined as the dynamic of a particle with a unit mass, which is subject to force
${\bf F}$. The ISG-method provides us an algorithmic and systematic solution processes for solving second-order ODEs of the form
equation \eqref{2.1}, in which the force function $\phi(t, q,\dot{q})=P/Q$ is a fractional function of polynomials $P$ and $Q$ of the variables
$t$, $q$, and $\dot{q}$ with coefficients in the set of complex numbers.
In this work, we use the ISG-method to solve second-order non-linear ODEs like \eqref{2.1},
in which the force function is as in those of \eqref{2.1}. This method can be developed to solve non-linear ODEs of higher orders and even for system of ODEs of any order, which are not our discussion here.
The general strategy of the ISG-method for solving second-order ODE \eqref{2.1} is to systematically and algorithmically extract a certain
number (equal to the order of the DE, which is two here) of the independent ``first integral'' (or first-order ``invariant'') such
as $I_1 (t, q,\dot{q})=c_1$ and $I_2 (t, q,\dot{q})=c_2$ associated to the DE \eqref{2.1}, where $I_1$ and $I_2$ are known functions of the variables $t$, $q$ and $\dot{q}$
and $c_i$'s~($i$=1,2) are constants on the solutions of the DE \eqref{2.1}.
When we obtain these two invariants, the solution of the DE \eqref{2.1} is actually prepared, because one can
look at the first-order ODEs, $I_1 (t, q,\dot{q})=c_1$ and $I_2 (t, q,\dot{q})=c_2$ as a system of algebraic equations in terms of
variables $t, q$ and constants of motion $c_1$ and $c_2$. It is easy to solve this system of equations.
To solve this system of algebraic equations, it is enough to remove the variable $\dot{q}$ between them. For this purpose, one can obtain from one of them, for example,
$I_1 (t, q,\dot{q})=c_1$, the variable $\dot{q}$ in terms of $t$, $q$ and the constant of motion $c_1$ as follows: $\dot{q} =\chi_1(t, q; c_1)$
where $\chi_1$ is a known function of $t$, $q$ and $c_1$.
Then, by replacing this function in the second equation
$I_2 (t, q,\dot{q})=c_2$, one obtains the following equation:
\begin{equation}
c_2 = I_2 \big(t, q,\chi_1(t, q; c_1)\big):=\Sigma(t, q; c_1), \label{2.3}
\end{equation}
where $\Sigma$ is a known function of variables $t$, $q$  and constant of motion $c_1$.
Equation \eqref{2.3} can be solved algebraically and obtained $q=\Pi(t; c_1, c_2 )$, where $\Pi$ is a known function of variable $t$ and constants of motion
$c_1$ and $c_2$. In this way, the solution of the problem is complete, and $q=\Pi(t; c_1, c_2 )$ derived by the ISG-method
is the general solution of the DE \eqref{2.1}.

Now, to get into the details of the ISG-method, how can one find two independent invariants $I_1 (t, q,\dot{q})=c_1$ and $I_2 (t, q,\dot{q})=c_2$?
As mentioned before, if the force function $\phi(t, q,\dot{q})$ in the DE \eqref{2.1} is a fractional function of polynomials like $P$ and $Q$ of variables
$t$, $q$ and $\dot{q}$ with coefficients in the set of complex numbers as $\phi(t, q,\dot{q})=P/Q$, then, to find these two invariants, one can use the following theorem.
\\\\
{\bf Theorem 1.} (Duarte's integral formula)  {\it Let $\ddot{q} = \phi(t, q, \dot{q})$ be the DE
of a particle with one degree of freedom in the configuration space $\mathbb{Q}=(q)$, where the force function
$\phi(t, q,\dot{q})$ is a fractional function of polynomials $P$ and $Q$ of variables $t$, $q$ and $\dot{q}$. If this second-order ODE admits a first integral as $I(t, q, \dot{q})=c$, then}
\begin{eqnarray} \label{2.4}
I(t, q, \dot{q}) = r_1+ r_2- \int \Big[R + \frac{\partial}{\partial \dot{q}}( r_1+ r_2) \Big] d \dot{q},
\end{eqnarray}
{\it where}
\begin{eqnarray} \label{2.5}
r_1 = \int R(\phi + S \dot{q})  d t,~~~~~r_2 =- \int \Big(R S + \frac{\partial r_1}{\partial q} \Big)  d q.
\end{eqnarray}
{\it In equations \eqref{2.4} and \eqref{2.5}, $S$ and $R$, which satisfy in the PS determining equations: }
\begin{eqnarray}
D[S]&=&-\phi_{_q} + S \phi_{_{\dot{q}}} +S^2, \label{2.6}\\
D[R]&=&-R(S + \phi_{_{\dot{q}}}), \label{2.7}\\
R_q&=& R_{_{\dot{q}}} S + R  S_{_{\dot{q}}}, \label{2.8}
\end{eqnarray}
{\it  are called the ``null form'' and the ``integrating factor'' of the DE \eqref{2.1}, respectively  }\cite{{ref.21},{ref.22},{ref.28}}.
{\it Here, $\phi_{_q}:=\frac{\partial \phi}{\partial q}, \phi_{_{\dot{q}}}:=\frac{\partial \phi}{\partial \dot{q}}$ and so on.}\\

In order to calculate the functionally independent invariants $I_1 (t, q,\dot{q})=c_1$ and $I_2 (t, q,\dot{q})=c_2$  associated to the DE \eqref{2.1},
we must first obtain the null form $S$ and integrating factor $R$, using the PS determining equations \eqref{2.6}-\eqref{2.8}.
Solving these differential equations for $S$ and $R$ is difficult, except for certain simple cases. Therefore, to obtain these
two basic functions in the PS procedure, we must resort to another symmetry methods. These methods are Lie
point symmetry, $\lambda$-symmetry and DPs.

If the calculation of functions $S$ and $R$ is not achieved by the PS determining equations, then one must do them indirectly. For this purpose, suppose \cite{ref.27}
\begin{eqnarray} \label{2.9}
&&\mathbf{\Phi}:\hspace{-1mm}\Bbb{R}^2 \rightarrow \Bbb{R}^2,\nonumber \\
&&(t, q)\mapsto (\overline{t},\overline{q})=\mathbf{\Phi} (t, q; \epsilon)=\big(T(t, q; \epsilon) , Q(t, q; \epsilon)\big),
\end{eqnarray}
with the transformation equations
\begin{eqnarray}\label{2.10}
&&{ t} \rightarrow {\bar t} =  t+\epsilon {\tau}(t, q)+ {\cal O} (\epsilon^2),\nonumber\\
&&q \rightarrow {\bar q} =  {q} + \epsilon {\xi}(t, {q}) + {\cal O} (\epsilon^2),~~\epsilon \ll 1,
\end{eqnarray}
be the group of one-parameter Lie point transformations
such that the DE \eqref{2.1}
under which remains invariant. In equations \eqref{2.10}, functions $\tau$ and $\xi$ defined by
\begin{eqnarray}
{\tau}(t, { q}) &=&\frac{\partial T(t, {q}; \epsilon)}{\partial \epsilon}{|_{_{\epsilon =0}}},\label{2.11}\\
{\xi}(t, {q}) &=&  \frac{\partial {Q}(t, {q}; \epsilon)}{\partial \epsilon}{|_{_{\epsilon =0}}},\label{2.12}
\end{eqnarray}
are called infinitesimals of the group of transformations $\mathbf{\Phi}$.
By these two functions, the infinitesimal generator vector of the group of transformations $\mathbf{\Phi}$ is defined as
\begin{eqnarray}\label{2.13}
{\bf X}= \tau(t, q) \frac{\partial}{\partial t}+\xi(t, q) \frac{\partial}{\partial q}.
\end{eqnarray}
In order to find the Lie point symmetries associated to the DE \eqref{2.1}, we must first obtain the infinitesimals $\tau$ and $\xi$.
For this purpose, we use the following theorem.
\\\\
{\bf Theorem 2.} (Lie's invariance condition) {\it  Suppose for a dynamical system with one degree of freedom, the second-order ODE
 $H(t, q, \dot{q}, \ddot{q}):=\ddot{q}-\phi(t, q, \dot{q})=0$ be the DE (or governing equation) of the particle in the configuration space,
 and let ${\bf X}=\tau(t,x) \partial_{_t}+\xi(t,q) \partial_{_q}$ be the infinitesimal generator of the one-parameter Lie group of transformations
\eqref{2.9} with transformation equations \eqref{2.10} acting on the $(t,q)$-space.
The group of transformations $\mathbf{\Phi}$ is admitted by the DE  \eqref{2.1}, if and only if the following condition is holds:}
\begin{eqnarray}\label{2.14}
{{{\bf X}{^{(2)}} H(t, q, \dot{q}, \ddot{q})}{\Big |}}_{H=0}=0,
\end{eqnarray}
{\it where}
\begin{eqnarray}\label{2.15}
{\bf X}{^{(2)}} :=\tau \frac{\partial}{\partial t} +\xi \frac{\partial}{\partial q}+ (\dot \xi -\dot q \dot \tau) \frac{\partial}{\partial \dot q}  +
(\ddot \xi -2 \ddot q \dot \tau -\dot q \ddot \tau ) \frac{\partial}{\partial \ddot q},
\end{eqnarray}
{\it is the 2th extended (prolonged) infinitesimal generator of the vector field
${\bf X}$. We note that equation \eqref{2.14} is called the
Lie's invariance condition which is the fundamental equation of the Lie symmetry analysis} \cite{{ref.27},{ref.29},{ref.30}}.

In the other words, the establishing equation \eqref{2.14} is a necessary and sufficient condition for the ODE $H=0$ to admit $\mathbf{\Phi}$
with infinitesimal generator $\bf X$ as a one-parameter Lie group of transformations.
By solving the Lie's invariance condition, which are in fact a system of second-order partial differential equations for the unknown functions $\tau$ and $\xi$,
all infinitesimal generators $\bf X$ associated to the ODE $H=0$ can be found.

Assume that the ODE $H=0$
has at least two Lie point symmetries and  ${\Phi_1}$ and ${\Phi_2}$ be arbitrarily two of them with the infinitesimal generators
${\bf X}_1=\tau_1 \partial_{_t}+\xi_1 \partial_{_q}$ and ${\bf X}_2=\tau_2 \partial_{_t}+\xi_2 \partial_{_q}$, respectively.
By the symmetries ${\bf X}_1$ and ${\bf X}_2$, we define the characteristics
$Q_1:=\xi_1 -\dot q \tau_1$ and $Q_2:=\xi_2 -\dot q \tau_2$, respectively.
Similarly, by the characteristics $Q_1$ and $Q_2$ we define
the functions $\lambda_i:=D[Q_i]/Q_i$, ($i$=1, 2).  It can be shown that $-\lambda_i$'s are solutions
of the determining equation \eqref{2.6}. Thus, we conclude that the null forms are $S_i=-\lambda_i$.
Now, by using the Darboux's eigenvalue equation\footnote{In this equation, $F$'s are DPs associated to the DE \eqref{2.1}, and the eigenvalue $K(t, q,\dot{q}) \big(\phi_{\dot{q}} (t, q,\dot{q})\big)$  is called the cofactor of $F$
when DE is an explicit function of  $t$ (when DE is not an explicit function of  $t$) and has degree at most 2.} \cite{{ref.21},{ref.27}}:
\begin{eqnarray}\label{2.16}
D[F] =
\begin{cases}
\phi_{_{\dot{q}}} F&~~~~ $if DE is not an explicit function of$~ t,\\
K F&~~~~$if DE is an explicit function of$~ t.\\
\end{cases}
\end{eqnarray}
one can obtain the DPs $F$ associated to the DE \eqref{2.1}. It can be shown that the ratio $Q/F$ is a solution of the determining equation \eqref{2.7}, and hence
$R=Q/F$. Therefore, two 2-tuples $(S_1 , R_1)$ and $(S_2 , R_2)$ are obtained as follow:
\begin{eqnarray}\label{2.17}
(S_1 , R_1)=(-\lambda_1 , \frac{Q_1}{F}), ~~~~~(S_2 , R_2)=(-\lambda_2 , \frac{Q_2}{F}).
\end{eqnarray}
Thus, by substituting each of these 2-tuples $(S_1 , R_1)$ and $(S_2 , R_2)$ in the Duarte's integral formula \eqref{2.4}
one can obtain the first integrals corresponding to each of these 2-tuples.
Certainly, these first integrals are independent of each other.
Therefore, they are what we needed to solve the DE \eqref{2.1}.
It should be noted if the DE \eqref{2.1} does not have any Lie point symmetry that can be used to calculate the null form $S$ and the integrating factor $R$,
then, one must use its generalized symmetries. The most suitable of these symmetries is $\lambda$-symmetry. Similar to Lie point symmetry,
$\lambda$-symmetry has also a symmetry condition. If this symmetry condition is used for the DE \eqref{2.1}, then,
the function $\lambda(t, q, \dot q)$ and its infinitesimals $\tau$ and $\xi$ are obtained. When the functions $\lambda$, $\tau$ and $\xi$ are known,
then the first integral can be calculated by a process which is given in Ref. \cite{ref.27}.

%%%%%%%%%%%%%%%%%%%%%%%%%%%%%%%%%%%%%%%%%%%%%%%%%%%%%%%%%%%%%%%%%%%%%%%%%%%%%%%%%%%%%%%%%%%%

\section{FLRW cosmological model in BD theory}
\label{Sec.III}

The purpose of this section is to study the FLRW cosmological model in the BD theory. Before proceeding to do this, let us give a short review of the theory of BD.

\subsection{A short review of the theory of BD}
\label{Sec.III.1}

The BD theory of gravity is a generalization of Einstein's theory of general relativity, which was formulated by Brans and Dicke \cite{{ref.3},{ref.4}}.
As mentioned earlier, this theory is consistent with Mach's principle (the stronger form) and relies less on the inherent properties of space.
In this theory, in addition to the tensor field that describes the geometry of space-time, a scalar field is also included.
Hence, BD theory is also called scalar-tensor theory. To obtain the field equations of this theory, let us consider Einstein-Hilbert action in the theory of GR
\begin{eqnarray}\label{3.1}
I_{_{EH}}=\int_{\cal M} \sqrt{-g} \Big(-\frac{\cal R}{16 \pi G} +{\mathcal{L}}_{_M} \Big) d^4x.
\end{eqnarray}
where ${\cal R}$ is the Ricci scalar associated to the four-dimensional space-time manifold ${\cal M}$ with the local coordinates $x^{\mu}=(x^{0},x^{1},x^{2},x^{3})$,
and ${\mathcal{L}}_{_{M}}={\mathcal{L}}_{_{M}}(g_{_{\mu \nu}}, \partial_{_{k}}g_{_{\mu \nu}})$ is the Lagrangian density of the ordinary matter and
$g:=det~g_{_{\mu \nu}}$. This action in the BD theory and in the ``Jordan frame'' is generalized as follows \cite{{ref.10},{ref.12}}:
\begin{eqnarray}\label{3.2}
I_{_{BD}}=\frac{1}{16\pi}\int_{\cal M} \sqrt{-g} \Big(\varphi {\cal R}-\frac{\omega}{\varphi}~g^{\mu \nu}~\nabla_{\mu} \varphi~\nabla_{\nu} \varphi-V(\varphi)
+16\pi~{\mathcal{L}}_{_{M}}\Big)  d^4x,
\end{eqnarray}
where $\varphi$ is the BD scalar field, $\omega$ is the adjustable BD coupling parameter, $V(\varphi)$ is the potential energy of the field $\varphi$ and
$\nabla_{\mu}$ denotes the covariant derivative operator in the space-time $x^{\mu}$.
Moreover, the BD scalar field $\varphi$ inversely proportional to the effective gravitational constant $G_{_{eff}}$ by the following relation
\begin{eqnarray}\label{3.3}
\varphi=\frac{1}{G_{eff}}~\frac{4+2\omega}{3+2\omega},
\end{eqnarray}
where $G_{eff}$ is equal to the constant $G$ in the limit when $\omega$ tends to infinity.
In addition, ${\mathcal{L}}_{M}$ as the Lagrangian density is minimally coupled to the BD scalar field $\varphi$.
In analogy to the GR theory, the law of conservation of energy-momentum of the ordinary matter $\nabla_{\mu}T^{\mu \nu}_{M}=0$
and the definition of the energy-momentum tensor (EMT) of the ordinary matter fields in the four-dimensional space-time is given by
\begin{eqnarray}\label{3.4}
T^{\mu \nu}_{M}:=\frac{2}{\sqrt{-g}}~\frac{\partial}{\partial g_{_{\mu \nu}}}~(\sqrt{-g}~\mathcal{L}_{_{M}}).
\end{eqnarray}
By varying the action \eqref{3.2} with respect to the scalar field $\varphi$ and metric  $g^{\mu \nu}$ one can get the following equations, respectively \cite{ref.31}
\begin{eqnarray}
&&\frac{2\omega}{\varphi}~\square \varphi+{\cal R}-\frac{\omega}{\varphi^{2}}~\nabla^{\alpha}\varphi~\nabla_{\alpha}\varphi-\frac{dV}{d\varphi}=0,\label{3.5.1}\\
&&{\cal R}_{\mu \nu}-\frac{1}{2}g_{\mu \nu}~{\cal R}=\frac{8\pi}{\varphi}~T_{M\mu \nu}
+\frac{\omega}{\varphi^{2}}(\nabla_{\mu}\varphi~\nabla_{\nu}\varphi-\frac{1}{2}g_{\mu \nu}~\nabla^{\alpha}\varphi~\nabla_{\alpha}\varphi)\nonumber\\
&&~~~~~~~~~~~~~~~~~~~~~~~~+\frac{1}{\varphi}(\nabla_{_{\mu}}\nabla_{_{\nu}}\varphi-g_{_{\mu \nu}}\square\varphi)-\frac{V}{2\varphi}g_{_{\mu \nu}},\label{3.6}
\end{eqnarray}
where ${\cal R}_{\mu \nu}$ is the Ricci tensor and $\square:=\nabla^{\alpha}~\nabla_{\alpha}$ is the covariant d'Alemberian operator of the metric.
Note that equation \eqref{3.5.1} is known as the ``modified Klein-Gordon equation'', and equation \eqref{3.6} is a generalization of the EFEs
 ${\cal R}_{\mu \nu}-(1/2)g_{\mu \nu}~{\cal R}+\Lambda g_{\mu \nu}=8\pi ~G ~T_{M \mu \nu}$. Now, by contracting on equation \eqref{3.6} one obtains
\begin{eqnarray}\label{3.7}
{\cal R}=-\frac{8 \pi T_{M}}{\varphi}+\frac{\omega}{\varphi^{2}}~\nabla^{\alpha}\varphi~\nabla_{\alpha}\varphi+\frac{3\square\varphi}{\varphi}+\frac{2V}{\varphi}.
\end{eqnarray}
In order to eliminate the Ricci scalar ${\cal R}$ from \eqref{3.5.1} one may insert \eqref{3.7} into \eqref{3.5.1}. It then results
\begin{eqnarray}\label{3.8}
\square\varphi=\frac{8\pi}{3+2\omega}~(8\pi T_{M}+\varphi\frac{dV}{d\varphi}-2V),
\end{eqnarray}
where $T_{M}:=g^{\mu \nu}~{T_{_M}}_{\mu \nu}$ is the trace of the energy-momentum tensor of the ordinary matter fields.
The equation \eqref{3.8} together with \eqref{3.6} form a system of coupled second-order non-linear differential equations which are called the general form of the BD equations.
Choosing the potential $V(\varphi)$ in the form $V(\varphi)=2\Lambda \varphi$, the field equations
\eqref{3.6} and \eqref{3.8} yield the following standard BD equations
\begin{eqnarray}
\square\varphi&=&\frac{8\pi}{3+2\omega}~T_{M}-\frac{2\Lambda\varphi}{3+2\omega},\label{3.9}\\
{\cal R}_{\mu \nu}-\frac{1}{2}g_{\mu \nu}~{\cal R}+\Lambda g_{\mu \nu}&=&\frac{8\pi}{\varphi}~T_{M\mu \nu}
+\frac{\omega}{\varphi^{2}}(\nabla_{\mu}\varphi~\nabla_{\nu}\varphi-\frac{1}{2}g_{\mu \nu}~\nabla^{\alpha}\varphi~\nabla_{\alpha}\varphi)\nonumber\\
&&+\frac{1}{\varphi}(\nabla_{\mu}\nabla_{\nu}-g_{\mu \nu}\square \varphi).\label{3.10}
\end{eqnarray}
This system of the field equations, without considering the cosmological constant $\Lambda$, were obtained for the first time by Brans and Dicke in 1961,
and known as the BD equations in the presence of the cosmological constant $\Lambda$.
These equations form the basis of the BD theory. In fact, BD theory with the BD equations as the field equations is an extended gravity theory of the ``scalar-tensor'' type,
in which the potential energy function sets to $V(\varphi)=2\Lambda \varphi$.
In BD theory, although the initial value of the BD coupling parameter was very small, its present value exceeds 500 in many cosmological models,
such that $\omega>500$ is in good agreement with experimental tests and general relativity predictions (for example, neutron stars, black holes, gravitational waves, binary pulsars, etc.) \cite{ref.32}.
The BD theory of gravity reduces the GR theory in the $\omega \rightarrow \infty$ limit \cite{{ref.33},{ref.34}}.

In the following, we will use the ISG-method to solve the DEs of the FLRW cosmological model in the framework of the BD theory.
In fact, we will obtain the cosmological solutions of the BD equations \eqref{3.9} and \eqref{3.10} by the ISG-method.

%%%%%%%%%%%%%%%%%%%%%%%%%%%%%%%%%%%%%%%%%%%%%%%%%%%%%%%%%%%%%%%%%%%%%%%%%%%%%%%%%%%%%%%%%%%%%%%%%%%%%%%%%%%%%%%%%%%%%%%%%%%%%%%%%%%%%%%%%%%%%%%%%%%%%%
\subsection{FLRW cosmological model in BD theory}
\label{Sec.III.2}
In this subsection, the FLRW cosmological model is briefly studied in the framework of the BD theory.
We consider a homogeneous and isotropic universe as a cosmological model for our study.
To a good approximation, this universe is described by the FLRW metric.
This metric in the coordinates $ x^\mu =(t,r,\theta,\varphi)$ is defined as
\begin{eqnarray}\label{3.11}
ds^2= -dt^2+a^{2}(t)\Big(\frac{dr^2}{1-kr^2}+r^2~d\theta ^{2}+r^2~sin^{2}\theta~ d\varphi ^{2}\Big),
\end{eqnarray}
where $a(t)$ is the scale factor, and the number $k$ can admits three values of $-1, 1, 0$ for spaces of negative, positive and zero curvature, which represents closed, open and flat universes, respectively \cite{{ref.35},{ref.36}}.
The Weyl's principle requires that the constituent material of a homogeneous and isotropic universe is a perfect fluid on large scales with high accuracy.
So, according to the Weyl's principle, the FLRW universe can be considered as a perfect fluid with a good approximation.
If the energy density of this fluid be $\rho c^2$ and its pressure $p$, then its normal barotropic equation of state is
\begin{eqnarray}\label{3.12}
p= w \rho c^2,
\end{eqnarray}
where $w$ is called the perfect fluid state parameter and $c$ is the velocity of light in vacuum.
The equation of state \eqref{3.12} includes most of the interesting states that are important in cosmology.
For example, for dust: $w=0$, for radiation: $w=1/3$, for false vacuum: $w=-1$, and for stiff fluid: $w=1$ \cite{{ref.37},{ref.38}}.
In the present work, a system of units is used in which the velocity of light in vacuum is set to unity. Furthermore, the components of the EMT of the fluid will be as follows:
\begin{eqnarray}\label{3.13}
T_{M\mu \nu}=(\rho+p)U_{\mu}~U_{\nu}+pg_{\mu \nu},
\end{eqnarray}
where $U^{\mu}=(1,0,0,0)$ is the co-moving 4-velocity. The trace of this tensor is given by
\begin{eqnarray}\label{3.14}
T_{M}=3p-\rho.
\end{eqnarray}
The non-vanishing components of the BDE \eqref{3.10} read
\begin{eqnarray}
\frac{3\dot{a}^2}{a^2}-\Lambda&=&\frac{8\pi}{\varphi}\rho+\frac{\omega}{2}\frac{\dot{\varphi}^2}{\varphi ^{2}}-3\frac{\dot{a}}{a}\frac{\dot{\varphi}}{\varphi},\label{3.15}\\
-\frac{2\ddot{a}}{a}-\frac{\dot{a}^{2}}{a^2}+\Lambda&=&
\frac{8\pi}{\varphi}p+\frac{\omega}{2}\frac{\dot{\varphi}^2}{\varphi ^{2}}+2\frac{\dot{a}}{a}\frac{\dot{\varphi}}{\varphi}+\frac{\ddot{\varphi}}{\varphi}.\label{3.16}
\end{eqnarray}
Also, using equation \eqref{3.14}, one concludes that equation \eqref{3.9} becomes
\begin{eqnarray}\label{3.17}
\frac{\ddot{\varphi}}{\varphi}+3\frac{\dot{a}}{a}\frac{\dot{\varphi}}{\varphi}=\frac{2\Lambda}{3+2\omega}+\frac{8\pi}{\varphi}\Big(\frac{\rho-3p}{3+2\omega} \Big).
\end{eqnarray}
Note that equation \eqref{3.17} is not an independent equation in
BD theory, because it can be obtained by using Bianchi identities $\nabla_{\mu}~G^{\mu \nu}=0$ and equations \eqref{3.15} and \eqref{3.16}.
So, we have only two independent equations to determine four unknown functions $a(t)$, $\varphi(t)$, $\rho(t)$ and $p(t)$.
As such this system of equations does not seem to have a unique solution.
Hence, to find these unknowns uniquely, two equations are needed.
One of these is the equation of state of the universe, \eqref{3.12}. To obtain the second equation we use the following Corollary.\\
\\
\textbf{Corollary1.}~{\it In the spatially flat $(k=0)$ FLRW cosmological model in the framework of the BD theory,
the necessary and sufficient condition for the deceleration parameter of the universe $q:=-a\ddot{a}/\dot{a}^2$ to be a constant,
is that a power-law between the cosmic scale factor $a(t)$ and the Brans-Dicke scalar field $\varphi(t)$ holds in the following form} \cite{{ref.35},{ref.36},{ref.37},{ref.38},{ref.39},{ref.40}}:
\begin{eqnarray}\label{3.18}
\varphi a^{n}=C,
\end{eqnarray}
{\it where $n$ and $C$ are the functions of the parameters $w$, $\omega$ and constants $G$ and $\Lambda$, that is, $C:=C(w,\omega,G,\Lambda)$ and $n:=n(w,\omega,G,\Lambda)$.}\\
\textbf{Proof.}~See Ref. \cite{ref.39}.

The history of this Corollary backs to the Dirac's hypothesis that states the constant $G$ is a variable quantity and should be related by a power-law relation such as
$G=\acute{C}a^n$ to the scale factor of the universe, where $\acute{C}$ and $n$ are some constants.
Since in the BD theory, the scalar field $\varphi$ is proportional to the inverse of the Newton's gravitational constant, that is, $\varphi=c_{_0}/G$,
(where $c_{_0}$ is the proportionality constant), combining this relation with $G={C'}a^n$ yields the power-law
in the form \eqref{3.18} in which $C:=c_{_0}/{C'}$.
By combining equations \eqref{3.15}-\eqref{3.17} with equation of state \eqref{3.12} and power-law relation \eqref{3.18}, we then obtain
\begin{eqnarray}
\frac{8\pi\rho}{C}a^n &=&\frac{\dot{a}^2}{a^2}\Big(\frac{6-\omega n^2-6n}{2}\Big)-\Lambda,\label{3.19}\\
\frac{8\pi w \rho}{C}a^n &=& (n-2)\frac{\ddot{a}}{a}+\frac{\dot{a}^2}{a^2}\Big[\frac{2(n-1)-(\omega+2)n^2}{2}\Big]+\Lambda,\label{3.20}\\
\frac{8\pi}{C}a^n(\rho-3 w \rho) &=&(3+2\omega)\Big[-n\frac{\ddot{a}}{a}+n(n-2)\frac{\dot{a}^2}{a^2}\Big]-2\Lambda.\label{3.21}
\end{eqnarray}
The combination of these three differential equations after performing some algebraic calculations leads to the
following non-linear second-order ODE for the scale factor $a(t)$
\begin{eqnarray}\label{3.22}
2(\omega n+3)\frac{\ddot{a}}{a}+(6+4n\omega-\omega n^2)\frac{\dot{a}^2}{a^2}=2\Lambda,~~~~~~~~~\omega n+3\neq 0.
\end{eqnarray}
This equation can be rewritten in the following form
\begin{eqnarray}\label{3.23}
\ddot{a}=\alpha\frac{\dot{a}^2}{a}+\beta a,
\end{eqnarray}
where the new parameters $\alpha$ and $\beta$ in terms of the old parameters $\omega$, $w$, $n$ and the constant $\Lambda$ are defined by
\begin{eqnarray}\label{3.24}
\alpha:=-\frac{6+4n\omega-\omega n^2}{2(\omega n+3)},~~~~~~~~~~~~~~~~~~\beta:=\frac{\Lambda}{\omega n+3}.
\end{eqnarray}
Therefore, the combination of BD equations with the fluid state equation of the universe and the power-law relation leads to a second-order ODE for the cosmic scale factor $a(t)$ in the form \eqref{3.23}.
In this work, we call this equation, the ``Friedman-Brans-Dicke equation'' (FBD-equation).
The FBD-equation alone describes the dynamics of spatially flat $(k=0)$ FLRW universe in the framework of the BD theory.
Equation \eqref{3.23} can be imagined as the equation of motion of a particle with a unit mass in one-dimension.
By defining the cosmic scale factor as the coordinate of this particle, i.e., $q(t):=a(t)$, equation \eqref{3.23} is the familiar form of Newton's second law
\begin{eqnarray}\label{3.25}
\ddot{q}=\phi(t,q,\dot{q}),
\end{eqnarray}
where
\begin{eqnarray}\label{3.26}
\phi(t,q,\dot{q}):=\alpha\frac{\dot{q}^2}{q}+\beta q,
\end{eqnarray}
is the component of the force function acting on the particle, and $\ddot{q}$ is the component of its acceleration vector.
Thus, to analyze the dynamics of cosmological model, it is enough to solve only the FBD-equation by the ISG-method. We will do this in the next section.
%%%%%%%%%%%%%%%%%%%%%%%%%%%%%%%%%%%%%%%%%%%%%%%%%%%%%%%%%%%%%%%%%%%%%%%%%%%%%%%%%%%%%%%%%%%%%%%%%%%%%%%%%%%%%%%%%%%%%%%%%%%%%%%%%%%%%%%%%%
\section{Solving the FBD-equation}
\label{Sec.IV}
In this section we employ the ISG-method to solve the FBD-equation \eqref{3.25}.
To this end, we use the following steps: \\
\\
(1)-We re-consider spatially flat $(k=0)$ FLRW cosmological model in the framework of BD theory with a dynamical system $S^{2}_{1}$.
As mentioned earlier, the governing equation of this system is $\ddot{q}=\phi(t,q,\dot{q})$, where $\phi(t,q,\dot{q})$ is the force function of a particle that moves in a $1$-dimensional mini-super space with configuration $\mathbb{Q}=(q)$ where $q:=a(t)$  is the coordinate of the particle.\\
\\
(2)-A one-parameter Lie group of point transformations ${\bf \Phi}(t,q;\varepsilon)$ with the transformation equations \eqref{2.10}
must be found (if there is a solution) such that the FBD-equation \eqref{3.25} remains invariant under this group of transformations.
For this purpose, by using Lie's invariance condition, a set of the PDEs should be extracted for the infinitesimals $\tau$ and $\xi$ of group ${\bf \Phi}$.
By simultaneously solving this system of PDEs, the infinitesimals $\tau$ and $\xi$ as functions of $t$ and $q$ should be obtained: $\tau=\tau(t,q)$, $\xi=\xi(t,q)$.
For our case, these functions are obtained as $\tau=\tau(t,q)=c_{1}$, $\xi=\xi(t,q)=c_{2}q$ where $c_{1}$ and $c_{1}$ are some arbitrary constants \cite{ref.27}.
Then, by using these results one must find all Lie point symmetry vectors of the transformations group ${\bf \Phi}$ such that for our case
these vectors are obtained to be
$\mathbf{X_{1}}=\partial_{t}$ and $\mathbf{X_{2}}=q\partial_{q}$ which are respectively the infinitesimal generator of the translation group along the time ${\bf \Phi}_{1}(t,q;\varepsilon)=(t+\varepsilon,q)$ and the scaling group ${\bf \Phi}_{2}(t,q;\varepsilon)=(t,e^{\varepsilon}q)$ \cite{ref.27}.\\
\\
(3)-To calculate the first integral corresponding
to the DE $\ddot{q}=\phi(t,q,\dot{q})$ of the particle in the configuration space $\mathbb{Q}=(q)$, which is equation \eqref{2.4} of Theorem 1,
the null form $S$ and integrating factor $R$ should be obtained by using the determining equations \eqref{2.6}-\eqref{2.8}.
It is difficult to solve these equations simultaneously, except in special simple cases.
Therefore, to obtain these two basic functions in the extended PS method, we must resort to other symmetry methods.
The most important other symmetry methods, that can be used here, are: $(a)$ Lie point symmetry, $(b)$ $\lambda$-symmetry, $(c)$ DPs.
Therefore, the functions $S$ and $R$ should be calculated indirectly by the symmetry methods $(a)$-$(c)$.\\
\\
(4)-Using the Lie point symmetry mentioned in section \ref{Sec.II},
the characteristic $Q(t,q,\dot{q}):=\xi-\dot{q}\tau$ and then $\lambda(t,q,\dot{q}):=D[Q]/Q$ must be defined.
It can be shown that $-D[Q]/Q$ is a solution of the determining equation \eqref{2.6}, in such a way that $S(t,q,\dot{q})=-\lambda(t,q,\dot{q})$.\\
\\
(5)-By using Darboux's eigenvalue equation \eqref{2.16} \cite{ref.21},
the DP ${F}(t,q,\dot{q})$ and the eigenvalue $K(t,q,\dot{q})(\phi_{\dot{q}}(t,q,\dot{q}))$ corresponding to this polynomial should be obtained.
It can be shown that the ratio $Q/F$ is a general solution of the determining equation \eqref{2.7}.
So, $R(t,q,\dot{q})=Q/F$. In this way, the null form $S$ and the integrating factor $R$ are obtained indirectly by the Lie point symmetry method and DPs, respectively, without the need to solve their determining equations. \\

Let us turn into the main goal of this section which is nothing but calculating the first integrals $I_{1}$ and $I_{2}$ corresponding to the FBD-equation  \eqref{3.25}.
Since the FBD-equation admits two Lie point symmetries, both $\lambda$-symmetries associated to FBD-equation can be only calculated by using the relation $\lambda=D[Q]/Q$
without solving the $\lambda$-symmetry condition \cite{ref.27}.
Therefore, both $\lambda$-symmetries associated to the Lie symmetry vectors $\mathbf{X}_{1}=\partial_{t}$ and $\mathbf{X}_{2}=q\partial_{q}$ are, respectively,
\begin{eqnarray}
\lambda_{1}(t,q,\dot{q})&=&\frac{D[Q_{1}]}{Q_{1}}=\frac{D[-\dot{q}]}{-\dot{q}}=\alpha\frac{\dot{q}}{q}+\beta\frac{q}{\dot{q}},\label{4.1}\\
\lambda_{2}(t,q,\dot{q})&=&\frac{D[Q_{2}]}{Q_{2}}=\frac{D[q]}{q}=\frac{\dot{q}}{q}.\label{4.2}
\end{eqnarray}
It can be easily check that functions $-D[Q_i]/{Q_i}$ are solutions of the determining equations \eqref{2.6}.
Thus, for the FBD-equation, these functions are nothing but two null forms associated to the Lie symmetry vectors $\textbf{X}_{1}=\partial_{t}$ and $\textbf{X}_{2}=q\partial_{q}$:
\begin{eqnarray}
S_{1}(t,q,\dot{q})&=&-\lambda_{1}(t,q,\dot{q})=-\alpha\frac{\dot{q}}{q}-\beta\frac{q}{\dot{q}},\label{4.3}\\
S_{2}(t,q,\dot{q})&=&-\lambda_{2}(t,q,\dot{q})=-\frac{\dot{q}}{q}.\label{4.4}
\end{eqnarray}
On the other hand, the determining equation for DPs of the FBD-equation $\ddot{q}=\phi(t,q,\dot{q})$ is \eqref{2.16}.
As mentioned above, $Q/F$ is a solution of the determining equation \eqref{2.7} such that $R=Q/F$, and hence by considering $S=-D[Q]/Q$ we arrive at
\begin{eqnarray}\label{4.5}
D[\frac{Q}{F}]+\frac{Q}{F}\Big(-\frac{D[Q]}{Q}+\phi_{\dot{q}}\Big)=0.
\end{eqnarray}
Furthermore, the characteristics $Q=\xi-\dot{q}\tau$ obtained from the $\lambda$-symmetry and the
DPs $F$ obtained from solving of the Darboux eigenvalue equation associated to the FBD-equation gives us the integrating factor $R$ as $R=Q/F$.
Therefore, by using $\lambda$-symmetry and DPs without
solving the determining equations, we can obtain $2$-tuple $(S,R)$ as follows $(S,R)=(-D[Q]/Q,Q/F)$.
In the ISG-method, the basic quantities of the Lie point symmetry, extended PS method, $\lambda$-symmetry, and DPs read $\tau$, $\xi$, $\lambda$, $F$, $R$ and $S$.
These quantities are related to each other as shown in \cite{ref.27}.

Now, we consider one of the Lie symmetries, for example $\textbf{X}_{1}=\partial_{t}$.
This vector is the infinitesimal generator of the group of translation time: ${\bf \Phi}_{1}(t,q;\varepsilon)=(t+\varepsilon,q)$.
The characteristic corresponding to this symmetry vector is $Q_{1}=\xi_{1}-\dot{q}\tau_{1}=-\dot{q}$ and $\lambda$
associated to this characteristic is given by \eqref{4.1}. Therefore, the null form associated to the infinitesimal generator $ \textbf{X}_{1}=\partial_{t}$ is obtained from equation \eqref{4.3}.
We find the integrating factor $R_{1}$ corresponding to the null form $S_{1}$.
For this purpose, by using the characteristic $Q_{1}=-\dot{q}$ and by considering DP\footnote{According to Table 1 of Ref.  \cite{ref.27}, by the characteristics $Q_{1}(t,q,\dot{q})=-\dot{q}, Q_{2}(t,q,\dot{q})=q$
and DPs $F_{1}^{(D)}, F_{2}^{(D)}, \cdots, F_{6}^{(D)}$, two null forms $S_{i}:=-D[Q_{i}]/Q_{i}, (i=1, 2)$ and twelve integrating factors $R_{ij}:=Q{i}/F_{j}^{(D)}, (i=1,2, j=1,2,...,6)$ can be defined.} $F_{6}^{(D)}=q^{2\alpha}$, the integrating factor $R_{1}$ reads
\begin{eqnarray}\label{4.6}
R_{1}=\frac{Q_{1}}{F_{6}^{(D)}}=-\frac{\dot{q}}{q^{2\alpha}},
\end{eqnarray}
and $2$-tuple $(S_{1},R_{1})$ is then obtained to be of the form
\begin{eqnarray}
(S_{1},R_{1})= (-\alpha\frac{\dot{q}}{q}-\beta\frac{q}{\dot{q}}~,~-\frac{\dot{q}}{q^{2\alpha}}).\label{4.7}
\end{eqnarray}
By substituting this $2$-tuple in the Duarte's integral formula \eqref{2.4}, and then by calculating the integrals, one can get the first integral associated to the infinitesimal generator
$\mathbf{X_{1}}=\partial_{t}$, giving us
\begin{eqnarray}
I_{1}(q,\dot{q})&=&\frac{1}{2}\dot{q}^{2}~q^{-2\alpha}-\frac{\beta}{2(1-\alpha)}q^{-2\alpha+2}=c_{1},\label{4.8}
\end{eqnarray}
where $c_{1}$ is an arbitrary constant. In the same way, the $2$-tuple $(S_{2},R_{2})$ corresponding to the symmetry vector $\textbf{X}_{2}=q\partial_{q}$ can be obtained as follows:
\begin{eqnarray}
(S_{2},R_{2})&=& (-\lambda_2 , Q_2/F_{1}^{(D)})= \big(-\frac{\dot{q}}{q}~,~ \frac{q}{\frac{\beta}{\alpha-1} q^2+\dot {q}^2}\big).\label{4.9}\\
	I_{2}(t,q,\dot{q}) &=&
	\begin{cases}
		(\alpha-1)t+\sqrt{\frac{\alpha-1}{\beta}}\tan^{-1}~(\frac{q}{\dot{q}}\sqrt{\frac{\beta}{\alpha-1}})=c_{2},~ &~~ \frac{\beta}{\alpha-1}>0,~~~~~\\
		(\alpha-1)t+\frac{q}{\dot{q}}=c_{2},  ~ &~~ \frac{\beta}{\alpha-1}=0,\\
		(\alpha-1)t+\sqrt{|\frac{\alpha-1}{\beta}|}\tanh^{-1}~(\frac{q}{\dot{q}}\sqrt{|\frac{\beta}{\alpha-1}|})=c_{2},~ &~~ \frac{\beta}{\alpha-1}<0.\label{4.10}
	\end{cases}
\end{eqnarray}
According to the above results and as earlier mentioned, a set of twelve members of the null forms and integrating factors can be constructed, which we call the PS set:
\begin{eqnarray}
S_{PS}=\{(S_{i},R_{ij})\}_{i=1,2,j=1,...,6}=\{(S_{1},R_{11}),(S_{1},R_{12}),(S_{1},R_{13}),(S_{1},R_{14}),(S_{1},R_{15}),\nonumber\\
(S_{1},R_{16}),(S_{2},R_{21}),(S_{2},R_{22}),(S_{2},R_{23}),(S_{2},R_{24}),(S_{2},R_{25}),(S_{2},R_{26})\}, \label{4.11}
\end{eqnarray}
By using the formula \eqref{2.4}, a first integral can be constructed with each of the $2$-tuple of this set.
Therefore, for twelve first integrals we have
\begin{eqnarray}\label{4.12}
I_{i,ij}=r_{1;ij}+r_{2;ij}-\int \big[R_{ij}+\frac{\partial}{\partial\dot{q}}(r_{1;ij}+r_{2;ij})\big]d\dot{q}=c_{i,ij},~~~~~~i=1,2; j=1,...,6,
\end{eqnarray}
where $c_{i,ij}$'s are some constants on the solutions of the DE $\ddot{q}=\phi(t,q,\dot{q})$. Also, according to \eqref{2.5}, $r_{1;ij}$'s and $r_{2;ij}$'s  are defined as
\begin{eqnarray}
r_{1;ij}(t,q,\dot{q})&=&\int R_{ij}(\phi+\dot{q}S_{i})dt,\label{4.13}\\
r_{2;ij}(t,q,\dot{q})&=&-\int[R_{ij}S_{i}+\frac{\partial}{\partial_{q}}(r_{i;ij})]dq. \label{4.14}
\end{eqnarray}
The FBD-equation $\ddot{q}=\phi(t,q,\dot{q})$ admits all members of the PS set \eqref{4.10} as $2$-tuple $(S,R)$.
The members of $S_{PS}$ and their corresponding first integrals have been listed in Table 2 in Ref. \cite{ref.27}.
Among the first integrals of Table 2, only two of them, $I_{1,16}$ and $I_{2,21}$, which we denoted by $I_{1}$ and $I_{2}$, respectively, are independent and the rest are dependent on two first integrals $I_{1}$ and $I_{2}$.
In the PS set \eqref{4.11}, $(S_{1},R_{16})$ and $(S_{2},R_{21})$ are actually the $2$-tuples $(S_{1},R_{1})$, $(S_{2},R_{2})$, respectively,
and the first integrals associated to these $2$-tuples are $I_{1}=I_{1,16}$ and $I_{2}=I_{2,21}$, respectively.

As was mentioned earlier, the infinitesimal generator of the translation group is the symmetry vector $\textbf{X}_{1}=\partial_{t}$.
Therefore, the first integral associated to this symmetry vector gives the energy of the dynamical system $S_{1}^{2}$.
Thus, the invariant $I_{1}$ is the energy of the dynamical system.
Now, consider the independent first integrals \eqref{4.8} and \eqref{4.10}. We look at these two independent invariants as a system of algebraic equations.
To solve this system of equations we must remove the variable $\dot{q}$ among the equations of this system.
To do this, we find $\dot{q}$ from the equation $I_{2}(t,q,\dot{q})=c_{2}$, and then employ $I_{1}(q,\dot{q})=c_{1}$. Finally,
the general solutions of the FBD-equation, which include all three cases $c<0$, $c=0$ and $c>0$, are worked out
\begin{eqnarray}\label{4.15}
	q(t) =
	\begin{cases}
		\Big(\frac{2c_{1}}{|c|}\Big)^{\frac{1}{2(1-\alpha)}}~\sinh^{\frac{1}{1-\alpha}}\Big[(\alpha-1)\sqrt{|c|}(t+\acute{c}_{2})\Big],~~~~~~ &~~~~~~ c<0,~~~~~\\
		\Big( 2c_{1}(\alpha -1)^2\Big)^{\frac{1}{2(1-\alpha)}}\Big(t+{c'}_{2}\Big)^{\frac{1}{1-\alpha}}, ~~~~~~&~~~~~~ c=0,\\
		\Big(\frac{2c_{1}}{c} \Big)^{\frac{1}{2(1-\alpha)}}~\sin^{\frac{1}{1-\alpha}}\Big[(\alpha-1)\sqrt{c}(t+\acute{c}_{2})\Big],~~~~~~ &~~~~~~ c>0.
	\end{cases}
\end{eqnarray}
where $\alpha$ and $\beta$ are given by \eqref{3.24}. Moreover,
\begin{eqnarray}
c = \frac{\beta}{\alpha-1},~~~~~~~~~~~~~{c'}_{2}&:=&\frac{c_{2}}{1-{\alpha}}.\label{4.16}
\end{eqnarray}
By redefinition ${c'}_{1}:=(2c_{1})^{\frac{1}{2(1-\alpha)}}$, the general solutions \eqref{4.15} can be written as follows:
\begin{eqnarray}\label{4.17}
	q(t) =
	\begin{cases}
		c'_1 \Big({\dfrac{-2 \Lambda}{s}}\Big)^{^{-\frac{w n +3}{s}}}\sin ^{^{\frac{2(w n +3)}{s}}}\Big[\dfrac{1}{w n +3}\sqrt{\frac{-\Lambda s}{2}} (t+c'_2)\Big],~~~~~~ &~~~~~~ \Lambda<0,~~~~~\\
		c'_1 \Big(\dfrac{s}{2(w n +3)}\Big)^{^{{\frac{2(w n +3)}{s}}}} \Big(t+c'_2\Big)^{^{\frac{2(w n +3)}{s}}}, ~~~~~~&~~~~~~ \Lambda=0,\\
		c'_1 \Big({\dfrac{2 \Lambda}{s}}\Big)^{^{-\frac{w n +3}{s}}}\sinh ^{^{\frac{2(w n +3)}{s}}}\Big[\dfrac{1}{w n +3}\sqrt{\frac{\Lambda s}{2}} (t+c'_2)\Big],~~~~~~ &~~~~~~ \Lambda>0.
	\end{cases}
\end{eqnarray}
where $s:=12+6n\omega -\omega n^2$.
As we expected, since the FBD-equation is an ODE of the second-order, its general solution contains two integration constants ${c'}_{1}$ and ${c'}_{2}$.
These constants can be determined by using the initial conditions of the problem, i.e., $q(t=0)=q_{0}$ and $\dot{q}(t=0)=\dot{q}_{0}$.

%%%%%%%%%%%%%%%%%%%%%%%%%%%%%%%%%%%%%%%%%%%%%%%%%%%%%%%%%%%%%%%%%%%%%%%%%%%%%%%%%%%%%%%%%%%%%%%%%%%%%%%%
\section{ Completing the solutions}
\label{Sec.V}
Now, we show that the parameter $n$ in the power-law equation \eqref{3.18} with the parameter of state $w$ in equation \eqref{3.12}, are consistent, provided that there is a relationship between them.
For this purpose, subtracting both sides of equations \eqref{3.19} and \eqref{3.20} we obtain
\begin{eqnarray}\label{5.1}
\frac{8\pi a^{n}}{C}\rho(1-w)=\frac{\dot{a}^2}{a^{2}}\Big( \frac{6-\omega n^2-6n}{2}-\frac{2(n-1)-(\omega+2)n^{2}}{2}\Big)-(n-2)\frac{\ddot{a}}{a}-2\Lambda.
\end{eqnarray}
By taking two consecutive times of the total derivative with respect to cosmic time $t$ from equation \eqref{3.18} one obtains
\begin{eqnarray}
\frac{\dot{a}}{a} &=& -\frac{1}{n}\frac{\dot{\varphi}}{\varphi},\label{5.2}\\
\frac{\ddot{a}}{a} &=& \frac{n+1}{n^2}\frac{\dot{\varphi}^2}{\varphi^2}-\frac{1}{n}\frac{\ddot{\varphi}}{\varphi}.\label{5.3}
\end{eqnarray}
By substituting \eqref{3.18}, \eqref{5.2} and \eqref{5.3} into equation \eqref{5.1} and after a little simplification we arrive at
\begin{eqnarray}\label{5.4}
\ddot{\varphi}-\frac{3}{n}\frac{\dot{\varphi}^2}{\varphi}=\frac{2n}{n-2}\Lambda\varphi+8\pi\rho(1-w)\frac{n}{n-2}.
\end{eqnarray}
In addition, one may insert equations \eqref{3.12}, \eqref{3.18} and \eqref{5.2} into \eqref{3.17} to get
\begin{eqnarray}\label{5.5}
\ddot{\varphi}-\frac{3}{n}\frac{\dot{\varphi}^2}{\varphi}=\frac{2\Lambda \varphi}{3+2\omega}+\frac{8\pi}{3+2\omega}\rho (1-3w).
\end{eqnarray}
Comparing equations \eqref{5.5} and \eqref{5.4} yields the following equation:
\begin{eqnarray}\label{5.6}
	n=
	\begin{cases}
	   \frac{1-3w}{\omega(w-1)-1},~~~~~~~\Lambda=0,~w \in[-1,1],\\
                 \\
	-\frac{1}{\omega+1},~~~~~~~~~~~~~~\Lambda\neq 0,~w =0.
	\end{cases}
\end{eqnarray}
It can be seen that the parameter $n$ in the power-law equation \eqref{3.18} cannot admit any value, but only those values satisfying in \eqref{5.6}.
Thus, one may use \eqref{5.6} to write the general solution \eqref{4.17} in terms of the parameters $\omega$, $w$ and $\Lambda$, and hence
by changing $q(t) \rightarrow a(t)$ we get
\begin{eqnarray}\label{5.7}
	a(t)=
	\begin{cases}
	 {c'}_1\Big(-\frac{2\Lambda(\omega+1)^2}{(4+3\omega)(3+2\omega)}\Big)^{-\frac{\omega+1}{3\omega+4}}\sin^{\frac{2(\omega +1)}{3\omega+4}}
    \Big[\sqrt{-(\frac{3\omega+4}{2\omega+3})\frac{\Lambda}{2}}(t+{c'}_2)\Big],~~~\Lambda<0,\\
     \\
     {c'}_1 \Big(\frac{4+3\omega(1-w^2)}{2[\omega(1-w)+1]}\Big)^{\frac{2[\omega(1-w)+1]}{4+3\omega(1-w^2)}}
     \big(t+{c'}_2\big)^{\frac{2[\omega(1-w)+1]}{4+3\omega(1-w^2)}},~~~~~~~~~~~~~~~~~~~~~\Lambda=0,\\
                 \\
	 {c'}_1\Big(\frac{2\Lambda(\omega+1)^2}{(4+3\omega)(3+2\omega)}\Big)^{-\frac{\omega+1}{3\omega+4}}\sinh^{\frac{2(\omega +1)}{3\omega+4}}
    \Big[\sqrt{(\frac{3\omega+4}{2\omega+3})\frac{\Lambda}{2}}(t+{c'}_2)\Big],~~~~~~~~\Lambda>0.
	\end{cases}
\end{eqnarray}
According to \eqref{5.6}, the above equation is actually the solution of the BD equations for the
scale factor of the dust-dominated universe in both cases $\Lambda< 0$ and $\Lambda > 0$, while the solution of the case  $\Lambda= 0$ can be included the universe fulfilling
with the dust, radiation and false vacuum, as well as stiff fluid.
Having the cosmic scale factor $a(t)$, one can calculate the BD scalar field $\varphi (t)$.
Using equations \eqref{3.18}, \eqref{5.6} and \eqref{5.7} and simplifying them we then get
\begin{eqnarray}\label{5.8}
	\varphi(t)=
	\begin{cases}
	 C {{c'}_1}^{\frac{1}{\omega+1}}  \Big(-\frac{2\Lambda(\omega+1)^2}{(4+3\omega)(3+2\omega)}\Big)^{-\frac{1}{3\omega+4}} \sin^{\frac{2}{3\omega+4}}
     \Big[\sqrt{-(\frac{3\omega+4}{2\omega+3}) \frac{\Lambda}{2}} (t+{c'}_2)\Big],~~~~\Lambda<0,\\
     \\
    C {{c'}_1}^{\frac{3 w-1}{\omega(w-1)-1}} \Big(\frac{4+3\omega(1-w^2)}{2[\omega(1-w)+1]}\Big)^{\frac{2(1-3 w)}{4+3\omega(1-w^2)}}
     \big(t+{c'}_2\big)^{\frac{2(1-3 w)}{4+3\omega(1-w^2)}},~~~~~~~~~~~~~~~\Lambda=0,\\
                 \\
	  C {{c'}_1}^{\frac{1}{\omega+1}}  \Big(\frac{2\Lambda(\omega+1)^2}{(4+3\omega)(3+2\omega)}\Big)^{-\frac{1}{3\omega+4}} \sinh^{\frac{2}{3\omega+4}}
     \Big[\sqrt{(\frac{3\omega+4}{2\omega+3}) \frac{\Lambda}{2}} (t+{c'}_2)\Big],~~~~~~~~\Lambda>0.
	\end{cases}
\end{eqnarray}
In order to calculate the energy density of the universe with the equation of state $p=\omega\rho$, in all three
cases $\Lambda < 0$, $\Lambda = 0$, and $\Lambda > 0$ we use equations \eqref{3.19}, \eqref{5.6} and \eqref{5.7}.
By doing some algebraic calculations, we get the following solution for the energy density of the universe
\begin{eqnarray}\label{5.9}
	\rho(t)=
	\begin{cases}
	 (-\frac{\Lambda C}{8\pi}){{c'}_1}^{-\frac{1}{\omega+1}}
     \Big(-\frac{2\Lambda(\omega+1)^2}{(4+3\omega)(3+2\omega)}\Big)^{-\frac{1}{3\omega+4}} \sin^{-\frac{6(\omega +1)}{3\omega+4}}
    \Big[\sqrt{-(\frac{3\omega+4}{2\omega+3})\frac{\Lambda}{2}}(t+\acute{c}_2)\Big], ~~\Lambda<0,\\
     \\
     M(\omega,w)X(\omega,w)(t+{c'}_2)^{N(\omega,w)},~~~~~~~~~~~~~~~~~~~~~~~~~~~~~~~~~~~~~~~~~~~~~~~~~~~~~~\Lambda=0,\\
                 \\
	(\frac{\Lambda C}{8\pi}){{c'}_1}^{-\frac{1}{\omega+1}}
     \Big(\frac{2\Lambda(\omega+1)^2}{(4+3\omega)(3+2\omega)}\Big)^{-\frac{1}{3\omega+4}} \sinh^{-\frac{6(\omega +1)}{3\omega+4}}
    \Big[\sqrt{(\frac{3\omega+4}{2\omega+3})\frac{\Lambda}{2}}(t+\acute{c}_2)\Big],~~~~~~~~~\Lambda>0,
	\end{cases}
\end{eqnarray}
where $M$, $X$, and $N$ are the functions of the parameters $\omega$ and $w$ which are defined as
\begin{eqnarray}
M(\omega,w)&:=&\frac{C~{{c'}_1}^{\frac{3w-1}{\omega(w-1)-1}}}{4\pi}\Big(\frac{4+3\omega(1-w^2)}{2[\omega(1-w)+1]} \Big)^{\frac{2(1-3w)}{4+3\omega(1-w^2)}},\label{5.10}\\
X(\omega,w)&:=&\frac{6(w-1)^2~\omega^2-\omega[12(w -1)+(1-3w)(3w-5)]+12-18w}{[4+3\omega (1-w^2)]^2},\label{5.11}\\
N(\omega,w)&:=&\frac{2(1-3w)-8-6(1-w^2)\omega}{4+3\omega (1-w^2)}.\label{5.12}
\end{eqnarray}
As mentioned earlier, the cases $\Lambda < 0$ and $\Lambda > 0$ of solution \eqref{5.9} give the energy density for dust dominated universe only,
while the case $\Lambda = 0$ of this solution gives the energy density of the perfect fluid-filled universe with equation of state  \eqref{3.12}.
Accordingly and also using the equation of state $p=w \rho$ and energy density \eqref{5.9}, the pressure of the universe is worked out to be
\begin{eqnarray}\label{5.13}
	p(t)=
	\begin{cases}
	 0~~~~~~~~~~~~~~~~~~~~~~~~~~~~~~~~~~~~~~~~~~~~~~~~~~~~~~\Lambda<0,\\
     \\
    w  M(\omega,w)X(\omega,w)(t+{c'}_2)^{N(\omega,w)},~~~~~~~~~~~~\Lambda=0,~~w\in [-1,1]\\
                 \\
	 0~~~~~~~~~~~~~~~~~~~~~~~~~~~~~~~~~~~~~~~~~~~~~~~~~~~~~\Lambda>0.
	\end{cases}
\end{eqnarray}
According to this solution, the pressure of the dust universe for $\Lambda < 0$, $\Lambda = 0$ and $\Lambda > 0$ is zero as we expected.
Furthermore, for $\Lambda = 0$, this solution gives the pressure of a universe filled by a perfect fluid.
We call the solutions \eqref{5.7}, \eqref{5.8}, \eqref{5.9} and \eqref{5.13} in this article as cosmological solutions of the BD equations.
It should be noted that these solutions were previously obtained in less detail without the use of the concept of symmetry by one of the authors of this article in \cite{{ref.41},{ref.42}}.
Note that when the coupling constant $\omega$ tends to infinity in solutions \eqref{5.7}, \eqref{5.8}, \eqref{5.9} and \eqref{5.13}, one concludes that
they are actually the cosmological solutions of general relativity,
which can be obtained from solving EFEs in the presence of the cosmological constant for a universe filled with a perfect fluid with the equation of state $p=w \rho$.

Before closing this section, let us assume that the perfect fluid, filling the universe, is radiation. For $w=1/3$ and $\Lambda =0$ it follows from \eqref{5.6} that $n=0$.
Then, equations \eqref{3.19} and \eqref{3.20} are reduced to
\begin{eqnarray}
\frac{8 \pi \rho}{C}&=& 3\frac{\dot{a}^2}{a^2},\label{5.14}\\
\frac{8 \pi p}{C}&=& -2\frac{\ddot{a}}{a}-\frac{\dot{a}^2}{a^2}.\label{5.15}
\end{eqnarray}
Now we compare these equations with the case $\Lambda =0$ of the following Friedmann equations in the theory of general relativity
\begin{eqnarray}
8 \pi G \rho&=& 3\frac{\dot{a}^2}{a^2},\label{5.16}\\
8 \pi G \rho&=& -2\frac{\ddot{a}}{a}-\frac{\dot{a}^2}{a^2}.\label{5.17}
\end{eqnarray}
In order to be compatible with general relativity in this special case (for radiation in the absence of the $\Lambda$), the value of the constant $C$ should be
\begin{eqnarray}
C(w=\frac{1}{3}, \omega, G, \Lambda=0)=\frac{1}{G}.\label{5.18}
\end{eqnarray}
By using condition \eqref{5.18}, one can obtain the cosmological solutions of the BD equations for radiation in the absence of the $\Lambda$, giving us
\begin{eqnarray}
a(t)={c'}_1\sqrt{2(t+{c'}_2)},~~~~\varphi (t)=\frac{1}{G},~~~~p(t)=\frac{1}{32\pi G t^2},~~~~~~~~\rho (t)={3}p(t).\label{5.19}
\end{eqnarray}
Therefore, when the condition \eqref{5.18} holds, then not only general relativity and Brans-Dicke theories are compatible with each other, but also the solutions of both theories will be the same for radiation.
This result is indicative of the fact that for a universe full of radiation $w=1/3$ whose metric is FLRW flat $(k=0)$,
in the absence of the cosmological constant $\Lambda$, the coupling constant of the
BD scalar field $\varphi$ with the gravitational field $g_{\mu \nu}$ is at its lowest possible, so that the two fields $\varphi$ and $g_{\mu \nu}$ are completely separate and have no paining with other.
In other words, the curvature of space-time is very large compared to the curvature of the scalar field.

%%%%%%%%%%%%%%%%%%%%%%%%%%%%%%%%%%%%%%%%%%%%%%%%%%%%%%%%%%%%%%%%%%%%%%%%%%%%%%
\section{Special solutions of the FBD-equation}
\label{Sec.VI}
In order to make sure the correctness of cosmological solutions of the BD equations \eqref{5.7}, \eqref{5.8}, \eqref{5.9} and \eqref{5.13}, it is necessary to compare our solutions with
special solutions such as Nariai's solutions \cite{ref.43}, O'Hanlon-Tupper vacuum solutions \cite{ref.8} and the inflation solutions that have been obtained before.
It can be shown that when $w\in [0,1/3]$ and ${c'}_2 =0$, our solutions include the Nariai's solutions.
Nariai's solutions in the absence of the $\Lambda$, when the fluid constituting the universe is dust, include the BD dust solutions
\begin{eqnarray}
a(t)&=&{c'}_1 \Big(\frac{4+3\omega}{2(\omega+1)} \Big)^{\frac{2(\omega +1)}{4+3\omega}}~~(t+{c'}_2)^{\frac{2(\omega +1)}{4+3\omega}},\label{6.1}\\
\varphi(t)&=&C\Big({c'}_1 \frac{(4+3\omega)}{2(\omega+1)} \Big)^{\frac{2}{4+3\omega}}~~(t+{c'}_2)^{\frac{2}{4+3\omega}},\label{6.2}\\
\rho (t)&=&M(\omega,0)~X(\omega ,0)~~~(t+{c'}_2)^{\frac{-6(\omega +1)}{4+3\omega}},\label{6.3}\\
p(t)&=&0.\label{6.4}
\end{eqnarray}
where $\omega \neq -4/3,-1.$
To obtain the vacuum solution ($p=0$, $\rho =0$),  we reconsider equation \eqref{3.19} that for the vanishing energy density we have
\begin{eqnarray}\label{6.5}
3-3n-\frac{\omega}{2}n^2=0,
\end{eqnarray}
The solutions of this quadratic equation are:
\begin{eqnarray}\label{6.6}
n_{\pm}=\frac{-3\pm \sqrt{3(3+2\omega)}}{\omega}.
\end{eqnarray}
The solutions $a(t)$ and $\phi(t)$ obtained for these two values of $n$ are called the ``vacuum solutions'' of the BD equations.
By substituting \eqref{5.6} into equation \eqref{6.6} one may obtain the parameter of the equation of state $w$ in terms of the coupling parameter $\omega$, expressing
\begin{eqnarray}\label{6.7}
w_{\pm}=\frac{-(2\omega+3)\pm (\omega+1)\sqrt{3(3+2\omega)}}{\pm \omega \sqrt{3(3+2\omega)}}.
\end{eqnarray}
For these two values of $w_{\pm}$, the cosmological solutions of BD equations in the absence of the $\Lambda$ are worked out
\begin{eqnarray}
a_{\pm}(t)&=& a_{0}(1+b t)^{q_{\pm}},\label{6.8}\\
\varphi_{\pm}(t)&=&\varphi_{0}(1+b t)^{1-3q_{\pm}},\label{6.9}
\end{eqnarray}
where
\begin{eqnarray}\nonumber
q_{\pm}=\frac{\omega+1\pm \sqrt{\frac{2\omega+3}{3}}}{3\omega+4},
\end{eqnarray}
also, the constants $a_{0}$, $\varphi_{0}$ and $b$  are
\begin{eqnarray}\nonumber
a_{0}:=a(t=0)={c'}_1\Big(\frac{{c'}_2}{q_{\pm}} \Big)^{q_{\pm}},~~~~~~
\varphi_{0}:=\varphi(t=0)=\frac{1}{G} ~{ {c'}_1}^{\frac{1-3q_{\pm}}{q_{\pm}}} \Big(\frac{{c'}_2}{q_{\pm}} \Big)^{1-3q_{\pm}},~~~~~~b:=\frac{1}{{c'}_2},
\end{eqnarray}
we note that the range of parameter $\omega$ is $\omega >-3/2$ with $\omega \neq -4/3, 0$.
These solutions are in agreement with the solutions of those of \cite{ref.15}.
The special solutions \eqref{6.8} and \eqref{6.9} were first obtained by O'Hanlon and Tupper in 1972 and are usually called the vacuum solutions of O'Hanlon-Tupper \cite{ref.8}. These solutions are completely consistent with Chauvet's solutions \cite{ref.9}.

At the end of this section, let us examine the inflation solutions of BD equations.
For the state parameter $w=-1$, the universe was passing through the false vacuum era \cite{{ref.44},{ref.45},{ref.46}}.
We expect that the solutions obtained from the BD equations in this particular state to be inflationary.
To this end, we begin with the case  $\Lambda =0$ of the cosmological solutions \eqref{5.7}, \eqref{5.8} and \eqref{5.9} for $w=-1$, which
are, respectively, given by
\begin{eqnarray}
a(t)&=&{c'}_1\Big(\frac{1}{\omega+\frac{1}{2}} \Big)^{\omega+\frac{1}{2}}~~(t+{c'}_2)^{\omega+\frac{1}{2}},\label{6.10}\\
\varphi(t)&=&\frac{1}{G}~{c'}_1^{\frac{2}{\omega+\frac{1}{2}}}\Big(\frac{2}{2\omega+1} \Big)^2~~(t+{c'}_2)^2,\label{6.11}\\
\rho(t)&=&\frac{1}{8\pi G}~{c'}_1^{\frac{2}{\omega+\frac{1}{2}}}~~\frac{(2\omega+3)(6\omega+5)}{(2\omega+1)^2}:=\rho_{-1}.\label{6.12}
\end{eqnarray}
where $\omega\neq -1/2,-3/2,-5/6$.
The last one shows that the energy density of the universe in the false vacuum era is a constant.
Here, we have denoted this constant value by $\rho_{-1}$.
Now, it is necessary to express the integration constant ${c'}_1$ in equations \eqref{6.10} and \eqref{6.11} in terms of cosmic scale factor of the universe at the time of the Big Bang $t=0$, in which we denote by $a_0$.
For this purpose, by putting $t=0$ in cosmological solution \eqref{6.10} and according to the initial condition $a_0=a(t=0)$, we then get
\begin{eqnarray}\label{6.13}
{c'}_2=\frac{2\omega+1}{2}\Big(\frac{a_0}{{c'}_1} \Big)^{\frac{2}{2\omega +1}}.
\end{eqnarray}
By substituting \eqref{6.12} and \eqref{6.13} into solutions \eqref{6.10} and \eqref{6.11}, one can obtain
\begin{eqnarray}
a(t)&=& a_0(1+\chi t)^{\omega+\frac{1}{2}},\label{6.14}\\
\varphi(t)&=&\frac{1}{G}~a_0^\frac{2}{\omega+\frac{1}{2}}~(1+\chi t)^2,\label{6.15}
\end{eqnarray}
where
\begin{eqnarray}\label{6.16}
\chi^2:=\frac{32\pi G \rho_{-1}~{c'}_1^{-\frac{2}{\omega+\frac{1}{2}}}}{(6\omega+5)(2\omega+3)},
\end{eqnarray}
is a constant.
The cosmic scale factor \eqref{6.14} and the BD scalar field \eqref{6.15} are the inflation
solutions of the BD equations in the absence of the cosmological constant, respectively.
When the coupling parameter $\omega$ tends to infinity, for very small times $\chi t\ll 1$, the limits of the inflation solutions of the BD equations become
\begin{eqnarray}
\lim_{_{\omega\rightarrow\infty}}a(t)&=& a_0~\lim_{_{\omega\rightarrow\infty}}\Big(1+\frac{1}{\omega}\sqrt{\frac{8\pi G\rho_{-1}}{3}}t \Big)^\omega =a_0~e^{\sqrt{\frac{8\pi G\rho_{-1}}{3}}t},\label{6.17}\\
\lim_{_{\omega\rightarrow\infty}}\varphi(t)&=&\frac{1}{G}.\label{6.18}
\end{eqnarray}
It can be seen that in the limit $\omega \rightarrow\infty$, the cosmic scale factor $a(t)$ changes exponentially with cosmic time $t$, while
the BD scalar field $\varphi (t)$ is a constant.
Notice that solutions \eqref{6.17} and \eqref{6.18} are the same solutions which are obtained from the theory of the GR in the absence of the cosmological constant in the case where the equation of state of the universe is in the form $p=-\rho_{-1}$.
For the case that $\chi t\gg 1$, then the inflationary solutions \eqref{6.14} and \eqref{6.15} are as follows:
\begin{eqnarray}
a(t)&=&a_0~(\chi t)^{\omega+\frac{1}{2}},\label{6.19}\\
\varphi(t)&=&\varphi_0~ (\chi t)^2,\label{6.20}
\end{eqnarray}
where
\begin{eqnarray}
\varphi_0=\frac{1}{G}~a_0^{\frac{2}{\omega+\frac{1}{2}}}.\label{6.21}
\end{eqnarray}
For the universe in the false vacuum era $(p=-\rho_{_{-1}})$ and in the absence of cosmological constant $\Lambda$,
it follows from \eqref{5.6} that
\begin{eqnarray}
n_{|_{w=-1,\Lambda=0}}=-\frac{2}{\omega+\frac{1}{2}}.\label{6.22}
\end{eqnarray}
Then, by substituting \eqref{6.22} into the power-law equation \eqref{3.18} and by using the fact that $C=1/G$, we find
\begin{eqnarray}
\frac{1}{G}=C=\varphi(t) ~a^n(t)=\varphi(t=0)~ a^n(t=0)=\varphi_{0}~a_{0}^n =\varphi_{0} a_0^{-\frac{2}{\omega+\frac{1}{2}}}.\nonumber
\end{eqnarray}
which is nothing but equation \eqref{6.21}.
This issue can be a confirmation of the correctness of the inflationary
solutions obtained by solving the BD equations for the false vacuum universe in the absence of the cosmological constant.
%%%%%%%%%%%%%%%%%%%%%%%%%%%%%%%%%%%%%%%%%%%%%%%%%%%%%%%%%%%%%%%%%%%%%%%%%%%%%%%%%%%%%%%%%%%%%%%%%%%%%%%%%%%%%%%%%%%%%%%%%%%%%%%%%%%%%%%%%%%%%%
\section{Conclusions}
\label{Sec.VII}
In this work, we have focused on solving the BD equations for spatially flat $(k=0)$ FLRW
universe fulfilling a perfect fluid with the equation of state $p=\omega\rho$, $(-1\leq \omega \leq 1)$.
Using equations \eqref{3.12} and \eqref{3.18} we have solved
the BD equations in both cases of the absence and presence of the cosmological constant to obtain functions $a(t)$, $\varphi(t)$, $\rho(t)$ and $p(t)$.
As was shown, the constant $n$ in equation \eqref{3.18} and the state equation parameter $w$ in \eqref{3.12} are compatible with each other, provided that
the condition \eqref{5.6} is held.
Equations \eqref{3.12} and \eqref{3.18} helped us summarize all BD equations in one equation, \eqref{3.25}, which we called FBD-equation.
This equation alone could describe the dynamics of the spatially flat FLRW cosmological model in the framework of BD theory.
As we have seen, this cosmological model could be imagined as a $1$-dimensional dynamical system with the configuration space $\mathbb{Q}=(a)$ in which the governing equation of particle motion was given by $\mathbf{F}=\phi(t,a,\dot{a})\mathbf{\partial}_{a}$. As the first step to solve the FBD-equation we found the corresponding Lie point symmetries. Then, we showed that
the FBD-equation has two independent point symmetries with infinitesimal generators $\mathbf{X}_{1}=\partial_{t}$ and $\mathbf{X}_{2}=a\partial_{a}$.
Using DPs and $\lambda$-symmetries, we found two independent $2$-tuple $(S_{1},R_{1})$ (equation \eqref{4.7}) and $(S_{2},R_{2})$ (equation \eqref{4.9})
consisting of null form $S$ and integrating factor $R$.
Using these independent $2$-tuples and applying Duarte's integral formula \eqref{2.4},
we obtained the two independent invariants \eqref{4.8} and \eqref{4.10} which were associated to the infinitesimal generators
$\mathbf{X}_{1}$ and $\mathbf{X}_{2}$, respectively.
We showed that these two independent invariants, each of which was a first-order ODE for the cosmic scale factor $a(t)$ in terms of cosmic time $t$, could be viewed as a system of algebraic equations
for the unknown functions $a(t)$ and $\dot{a}(t)$.
By eliminating $\dot{a}(t)$ between the two equations of the system, we obtained the analytical solution \eqref{5.7} for the cosmological scale factor $a(t)$.
By employing this solution in the BD equations, we were able to obtain the BD scalar field $\varphi(t)$, the energy density of the universe $\rho(t)$
and its pressure $p(t)$, which were given by equations \eqref{5.8}, \eqref{5.9}, and \eqref{5.13}, respectively.
Not only these solutions gave the scale factor $a(t)$, the BD scalar field $\varphi(t)$, the energy density $\rho(t)$ and pressure of the universe $p(t)$ for the dust,
but also they gave us the more general state of a perfect fluid with equation of state $p=\omega\rho$, $(-1\leq \omega \leq 1)$.

As an interesting result, we have shown that when the coupling parameter $\omega$ tends to infinity, the cosmological solutions of GR theory can be obtained from
the analytical solutions of BD equations \eqref{5.7}, \eqref{5.8}, \eqref{5.9} and \eqref{5.13};
this can be a confirmation of $\lim_{\omega\rightarrow \infty}$ BD~=~GR.
Note that our cosmological solutions are rich, so that they include many well-known special solutions that were previously obtained by other methods.
For example, the solutions presented by Uehara and Kim \cite{ref.12} are special cases of our solutions, when
the cosmological constant $\Lambda$ is present for the dust dominated universe.
In addition, as shown, the case $\Lambda=0$ of our solutions include the following famous solutions: Nariai's solutions $(w\in[0,1/3])$, BD dust solutions $(w =0)$, radiation solutions $(w =1/3)$,
inflationary solutions $(w=-1)$, O'Hanlon-Tupper vacuum solutions $(w=w_{\pm})$ and general relativity solutions $(\omega\rightarrow\infty)$.

\subsection*{Acknowledgements}

This work has been supported by the research vice chancellor of Azarbaijan Shahid Madani University under research fund No. 1402/917.
%%%%%%%%%%%%%%%%%%%%%%%%%%%%%%%%%%%%%%%%%%%%%%%%%%%%%%%%%%%%%%%%%%%%%%%%%%%%%

\end{document}